\documentclass[12pt,a4paper]{article}
\usepackage{graphicx} 
\usepackage{fullpage}

\title{\textbf{Study of particle production from quark and gluon jets in proton-proton collisions}}
\author{Sona Pochybova$^{1,2}$\\
		\small{sona.pochybova@cern.ch}\\
		\small{\textit{$^1$ MTA KFKI RMKI, Konkoly-Thege Miklos 29-33, Budapest, Hungary}}\\
		\small{\textit{$^2$ E\"otv\"os University, P\'azm\'any P\'eter s\'et\'any 1/A, 
H-1117 Budapest, Hungary}}
		}

\begin{document}
\maketitle
\abstract{
We investigate whether and how different fragmentation properties of quarks and gluons affect identified particle spectra.   
We present a systematic study of $\pi$, $K$ and $p$ production in minimum bias (inelastic, non-diffractive), two- and three-jet events at RHIC, Tevatron and LHC energies. Through the study of two- and three-jet events and various jet-production channels we can directly access the fragmentation properties of quark and gluon jets. We present MC estimate for the contribution of quark and gluon jets to individual particle species spectra, that can be compared to experimental results and test our current knowledge of the physics behind particle production inside jets.
}

\section{Introduction}
\label{sec:Sec1}
Jets are produced in hard scatterings of colliding particles. Emerging from the early stages of collisions they are ideal tools to study final states, hadronisation processes and hadron production. Such questions can be addressed through investigation of fragmentation properties of quark and gluon jets in different event shapes (2- or 3- jet) and jet-production channels.

In the following, we show how gluon contribution to hadron spectra changes with collision energy, moreover, that this contribution is dominant for protons. Further, in the case of 2- and 3-jet events, we show that extra hard gluon radiation has an effect of 20\% - 40\% to the relative proton production. The presented MC analysis is an extension of multiple jet studies (\cite{Abreu:1995hp} - \cite{Acosta:2004yy}), with the perspective of implication in the future experiments at LHC, which offer high momentum particle identification capabilities.

The data were simulated using PYTHIA event generator \cite{Sjostrand:2006za} with the settings of P0 tune \cite{Skands:2008}. Three types of data-sets, were created for each collision energy - $\mathrm{200\ GeV}$, $\mathrm{1800\ GeV}$ and $\mathrm{7\ TeV}$ including pure gluon (GG), pure quark (QQ) and mixed (QG) jet event production. Every sample contains 300 000 events. Separating  the production channels, we were able to see the parton-type effect on single hadron spectra. Additionally, as a reference, three respective minimum bias (MB) samples were generated, each containing 1M events.

In order to study particle production in different event shapes, we selected 2- and 3-jet-like events based on the so called thrust variable, $T$ \cite{Sjostrand:2006za}. Events with $T$ smaller than $0.9$ were treated as 3-jet-like, events with higher values as 2-jet-like. This separation was earlier proposed for multi-jet analysis at LEP, see e.g Ref. \cite{Barreiro:1985av}.
The effects of GG, QQ and QG production channels where studied via $p+\bar{p}/K^{+}+K^{-}$ (p/K) and $p+\bar{p}/\pi^{+}+\pi^{-}$ (p/$\pi$) ratios.
\section{Particle ratios in selected production channels}
\label{sec:Sec2}
\begin{figure}[ht]
\begin{center}
\includegraphics[width=150mm,height=90mm]{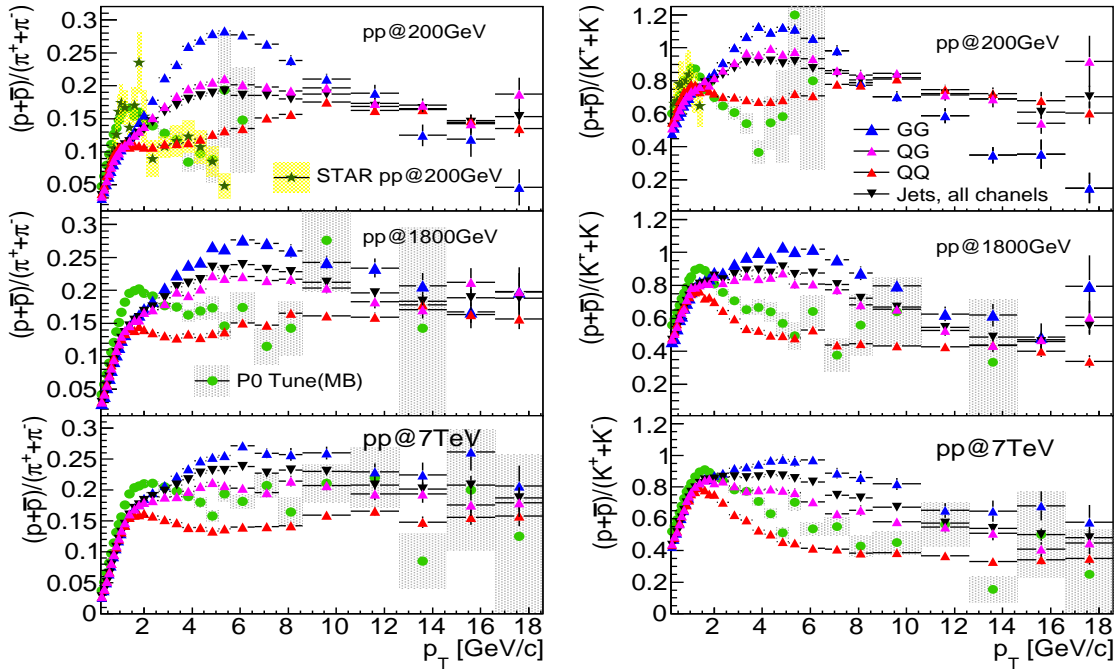} 
\end{center}
\caption{p/$\pi$ (left panels) and p/K (right panels) for various production channels; GG (blue triangles), QG (magenta triangles), QQ (red triangles) compared to all production channels (black triangles) and MB (green circles). At $\mathrm{\sqrt{s_{NN}}=200\ GeV}$ ratio is compared to STAR data \cite{STAR:2006}.}
\label{fig:dSigmaPt}
\end{figure}

At first we examined the separate contributions of individual jet-production channels to the whole spectra (see Table \ref{tab:table1}) and to the p/$\pi$ and p/K ratios (Figure \ref{fig:dSigmaPt}). 

\begin{table}
\caption{Integrated fraction of jet events in MB production (second column) and fraction of various production channels contained in a jet production.}

\begin{center}
\begin{tabular}{l || l l l l}
\hline
$\sqrt{s_{NN}}$ [TeV] & Jet/MB & GG/Jet & QQ/Jet & QG/Jet \\
\hline
$0.2$ & $0.1 \%$  & $17.7 \%$ & $27.3 \%$ & $55 \%$ \\
$1.8$ & $34.2 \%$ & $49.7 \%$ & $7.6 \%$  & $42.7 \%$\\
$7$   & $95 \%$   & $60 \%$   & $5.3 \%$  & $34.7 \%$\\
\hline
\end{tabular}
\end{center}
\label{tab:table1}
\end{table}

In the case of jet sample generation, we have taken into account, that the cross section for jet production varies strongly with jet energy. For this reason, the jet sample was generated in three $p_{T}^{hard}$ bins ($p_{T}^{hard}=\{15-50\ \mathrm{GeV/c}, 50-100\ \mathrm{GeV/c}, 100\ \mathrm{GeV/c} \leq \}$), where $p_{T}^{hard}$ is the transverse momentum of the partons in the rest frame of the interaction. After proper cross section scaling, the partial $p_{T}^{hard}$ samples where merged into one.

We can see, that the fraction of jet events contributing to the MB spectra rises with energy, furthermore, jet sample becomes gluon dominated. 

For both p/$\pi$ and p/K, in the $2-6\ \mathrm{GeV/c}$ region, the highest value of the ratio is achieved for GG events and the ratio for all channels rises towards the GG value with collision energy (Figure \ref{fig:dSigmaPt}). Specifically, p/$\pi$ ratio reaches values between $\approx 0.25-0.3$ and p/K ratio rises up to $\approx 1$. This behaviour can be connected to how protons are formed within the popcorn fragmentation model in PYTHIA \cite{Sjostrand:2006za}, \cite{Andersson:1983ia}. Observing the ratios experimentally can contribute to further understanding of particle production mechanisms.

The MB points seem to prefer the region between QG and QQ values, especially in the region above $\mathrm{3\ GeV/c}$, where hard scattering becomes important. In this sample we also observe a rising trend towards higher collision energies.   

\begin{figure}[ht]
\begin{center}
\includegraphics[width=150mm,height=50mm]{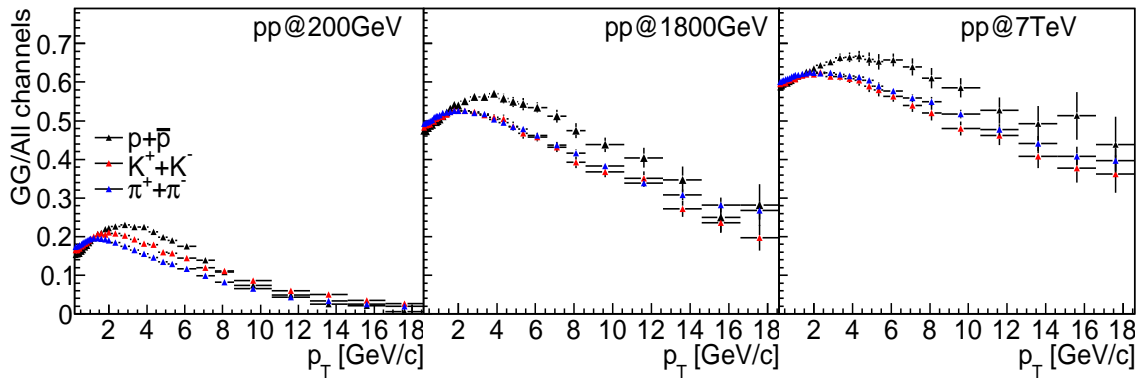}
\end{center}
\caption{GG production channel contribution to all production channels for individual particle spectra.}
\label{fig:GG} 
\end{figure}

In Figure \ref{fig:GG} we plotted the GG channel contribution to the jet spectra for individual hadrons as it changes with hadron momentum and collision energy. The contribution weakens with momentum and rises with collision energy (also see Table \ref{tab:table1}). The highest contribution belongs to protons, which complements Figure \ref{fig:dSigmaPt}.

\section{Particle ratios in 2- and 3-jet like events}
\label{sec:Sec3}
\begin{figure}[ht]
\begin{center}
\includegraphics[width=100mm,height=70mm]{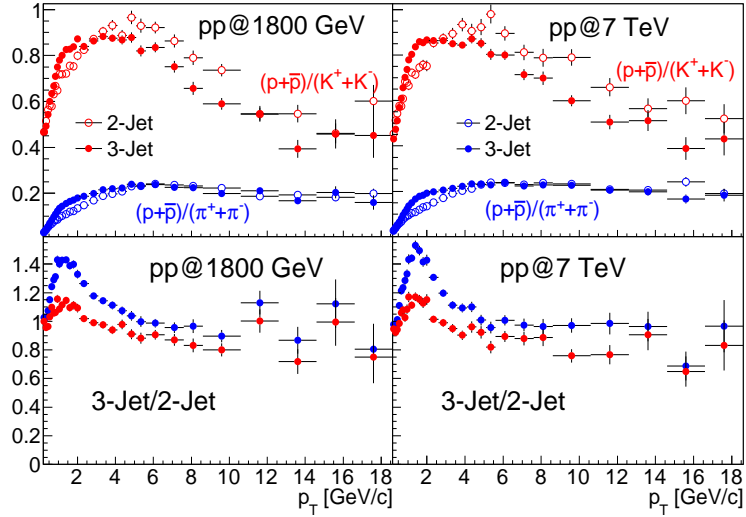}
\end{center}
\caption{Top panels: p/$\pi$ (blue) and  p/K (red) ratios for all production channels spectrum. Bottom panels: (3-jet ratio)/(2-jet ratio).}
\label{fig:RatioProtons} 

\end{figure}

In the second part of our investigation, we focused on how the studied ratios are changing in the presence of extra hard gluon radiation, i.e in 3-jet events.  
In Figure \ref{fig:RatioProtons} we plot the p/K and p/$\pi$ ratios for 2- and 3-jet like events. The extra hard gluon radiation, present in 3-jet like events, causes an excess in proton spectrum w.r.t pion or kaon. The effect being stronger for p/$\pi$. This is true for regions below $\mathrm{6\ GeV/c}$, where increase is up to $\approx 40\%$ for p/$\pi$ and up to $\approx 20\%$ for p/K. The enhancement, caused by hard gluon radiation, in the individual spectra originates, as mentioned in section \ref{sec:Sec2}, in the fragmentation model used within PYTHIA.

Above $\mathrm{6\ GeV/c}$, the differences between 2- and 3-jet like events vanish. This is consistent with Figure \ref{fig:GG} in the high $p_{T}$ region. The gluon contribution varies between protons as baryons, and pions and kaons as mesons. 

\section{Discussion}
\label{sec:Sec4}
We presented a MC study of particles originating from hard scatterings. Through the investigation of p/K and p/$\pi$ ratios in different jet-production channels we were able to see if and how the differences in the ratio values reflect the different fragmentation properties of quarks and gluons. Generally, we observed that gluon contribution to jet spectra rises with energy (see Table \ref{tab:table1}), and the gluon contribution to the spectra is highest for protons (Figure \ref{fig:dSigmaPt}, Figure \ref{fig:GG}).

When comparing 2- and 3-jet like events, we saw that additional hard gluon radiation, which is present in the 3-jet case, causes an enhancement in proton spectrum w.r.t pion (up to $\approx$ 40\%) as well as kaon (up to $\approx$ 20\%) (Figure \ref{fig:RatioProtons}).  

To conclude, the differences in particle production in the individual production channels (GG, QG and QQ) as well as 2- and 3-jet events are present in the mid-$p_{T}$ region ($\mathrm{2-6\ GeV/c}$) and are directly connected to the fragmentation model used in PYTHIA. Further they are related to the way how "soft" and "hard" parts of the event interplay. In this sense, the presented study suggests methods to investigate fragmentation and coalescence in an experimentally interesting momentum region for this kind of analysis.

\section*{Acknowledgements}
I would like to thank Peter L\'evai, Gergely G. Barnaf\" oldi and Levente Moln\'ar for fruitful discussions on the presented topic. The work has been funded by OTKA NK77816, PD73596 and E\"otv\"os University, Budapest.

\end{document}